\documentclass[12pt]{article}

\usepackage{epsfig}

\begin{document}

\title{\Large \bf Signals of the QGP phase transition - a view
from microscopic transport models }
\author{\large E.L. Bratkovskaya \bigskip \\
{\it  Frankfurt Institute for Advanced Studies,}\\
 {\it Ruth-Moufang-Str. 1,
 60438 Frankfurt am Main,
 Germany}  }

\maketitle

%{\large

\begin{center}
{\bf Abstract}\\
\medskip
In this contribution the results from various transport
models on different observables - considered as possible
 signals of the phase transition from hadronic matter to the
quark-gluon plasma (QGP) - are briefly reviewed.
\end{center}

\section{Introduction} \label{s1}
The phase transition from partonic degrees of freedom (quarks and
gluons) to interacting hadrons is a central topic of modern high-energy
physics. In order to understand the dynamics and relevant scales of
this transition laboratory experiments under controlled conditions are
presently performed with ultra-relativistic nucleus-nucleus collisions.
Hadronic spectra and relative hadron abundancies from these experiments
reflect  important aspects of the dynamics in the hot and dense zone
formed in the early phase of the reaction.

Estimates based on the Bjorken formula \cite{bjorken} for the
energy density achieved in central Au+Au collisions suggest that
the critical energy density for the formation of a quark-gluon
plasma (QGP) is by far exceeded during a few fm/c in the initial
phase of Au+Au collisions at Relativistic Heavy-Ion Collider
(RHIC) energies, but sufficient energy densities ($\sim$ 0.7-1
GeV/fm$^3$ \cite{Karsch}) might already be achieved at Alternating
Gradient Synchrotron (AGS) energies of $\sim$ 10 $A\cdot$GeV
\cite{exita}. More recently, lattice QCD calculations at finite
temperature and quark chemical potential $\mu_q$ \cite{Fodor} show
a rapid increase of the thermodynamic pressure $P$ with
temperature above the critical temperature $T_c$ for a phase
transition to the QGP. The crucial question is, however, at what
bombarding energies the conditions  for the phase transition are
fulfilled. Thus, it is very important to perform an 'energy scan'
of different observables in order to find an 'anomalous' behavior
that might be attributed to a phase transition.

In addition to the strong interactions in the initial stage of the
reaction - attributed to the QGP - there are also strong
(pre-)hadronic interactions after/during the  hadronization phase.
Thus it becomes very important to know the impact of such (pre-)
hadronic interactions on the final observables. The relevant
information on this issue can be provided by microscopic transport
models based on a nonequilibrium description of the nuclear
dynamics \cite{HORST}.

In this contribution I present the compilation of HSD results on
two of the possible signals of the phase transition: strangeness
and charm. The HSD (Hadron-String-Dynamics) transport approach
\cite{Geiss,Cass99} employs hadronic and string degrees of freedom
and takes into account the formation and multiple rescattering of
hadrons; it thus dynamically describes the generation of pressure
in the early phase - dominated by strings - and the hadronic
expansion phase. The HSD transport approach is matched to
reproduce the nucleon-nucleon, meson-nucleon and meson-meson cross
section data in a wide kinematic range. It also provides a good
description of particle production in p+A reactions \cite{Sibbi}
as well electroproduction of hadrons off nuclei \cite{Falter}. In
order to obtain a model independent conclusion, we also address
the results from the UrQMD model \cite{URQMD1,URQMD2} which has
similar underlying concepts as HSD but differs in the actual
realizations.

\section{Strangeness signals of the QGP} \label{s2}

As has been proposed in 1982 by Rafelski and M\"uller
\cite{Rafelski1} the strangeness degree of freedom might play an
important role in distinguishing hadronic and partonic dynamics.
In 1999  Ga\'zdzicki and Gorenstein \cite{SMES} - within the
statistical model - have predicted  experimental observables which
should show an anomalous behaviour at the phase transition: the
'kink' -- an enhancement of pion production in central Au+Au
(Pb+Pb) collisions relative to scaled $pp$ collisions; the 'horn'
-- a sharp maximum in the $K^+/\pi^+$ ratio at 20 to 30
A$\cdot$GeV; the 'step' -- an approximately constant slope of
$K^\pm$ spectra  starting from 20 to 30 A$\cdot$GeV. Indeed, such
"anomalies" have been observed experimentally by the NA49
Collaboration \cite{NA49_new,NA49_T}.

\begin{figure}[t]
\begin{minipage}[l]{8 cm}
\epsfig{file=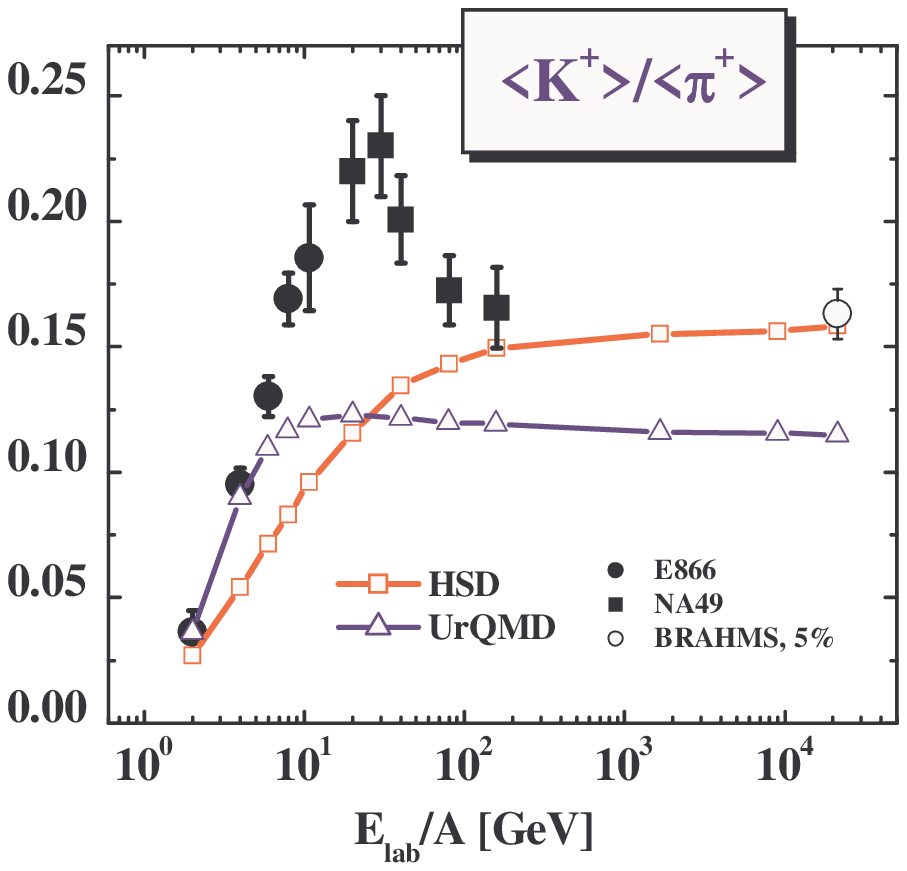,width=7cm}
\end{minipage}
\hfill\begin{minipage}[l]{8 cm}
\epsfig{file=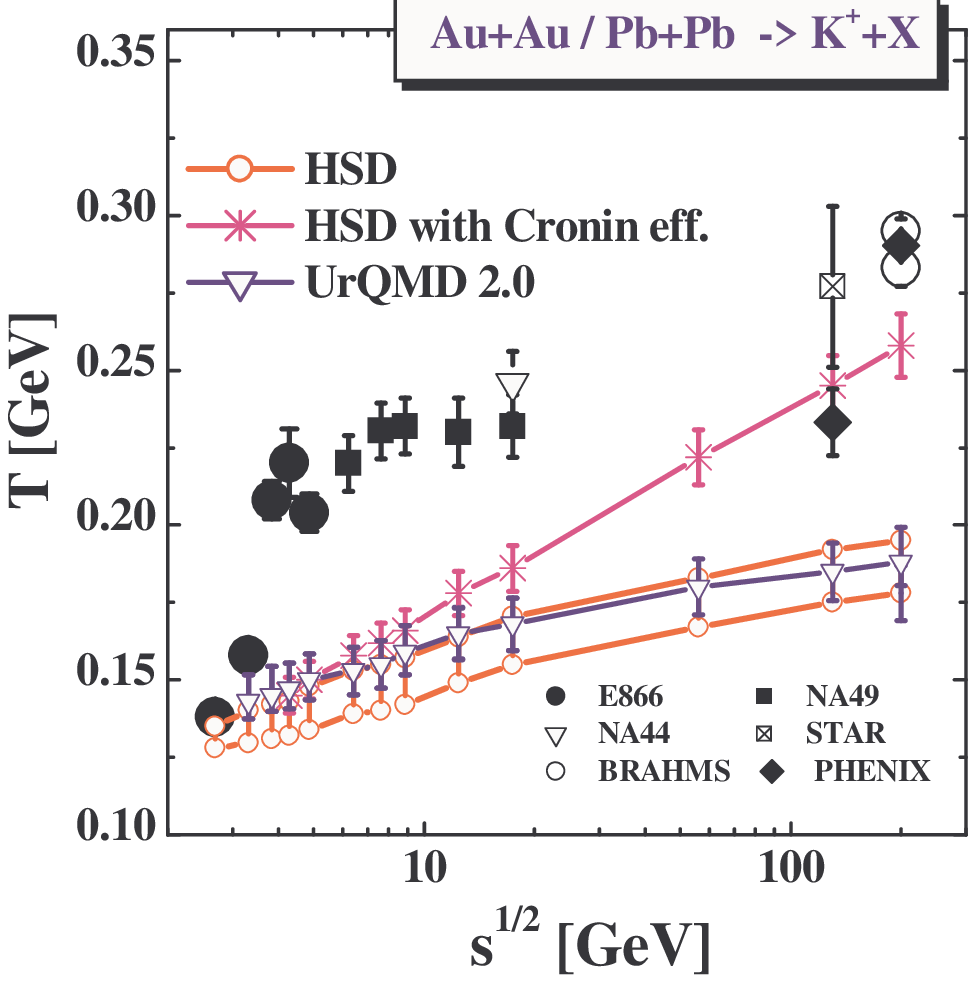,width=6.3cm}
\end{minipage}
\caption{Excitation function of the $K^+/\pi^+$ ratio (l.h.s.) and
inverse slope parameter for $K^+$ (r.h.s.) from central Au+Au (AGS
and RHIC) or Pb+Pb (SPS) collisions.  The solid lines with open
squares show the results from HSD whereas the dashed lines with
open triangles indicate the UrQMD calculations. The solid lines
with stars correspond to HSD calculations including 'Cronin'
initial state enhancement.} \label{Fig1ab}
\end{figure}

In Refs. \cite{Weber02,Brat03PRL,Bratnew} we have investigated the
hadron production as well as transverse hadron spectra in
nucleus-nucleus collisions from 2 $A\cdot$GeV to 21.3 $A\cdot$TeV
within the independent transport approaches UrQMD and HSD.  The
comparison to experimental data demonstrates that both approaches
agree quite well with each other and with the experimental data on
hadron production. The enhancement of pion production in central
Au+Au (Pb+Pb) collisions relative to scaled $pp$ collisions (the
'kink') is well described by both approaches without involving any
phase transition. However, the maximum in the $K^+/\pi^+$ ratio at
20 to 30 A$\cdot$GeV (the 'horn') is missed by $\sim$ 40\%
\cite{Weber02,Bratnew} -- cf. Fig. \ref{Fig1ab} (l.h.s.). A
comparison to the transverse mass spectra from $pp$ and C+C (or
Si+Si) reactions shows the reliability of the transport models for
light systems \cite{Brat03PRL}. For central Au+Au (Pb+Pb)
collisions at bombarding energies above $\sim$ 5 A$\cdot$GeV,
however, the measured $K^{\pm}$ $m_{T}$-spectra have a larger
inverse slope parameter than expected from the calculations. The
approximately constant slope of $K^\pm$ spectra at SPS (the
'step') is not reproduced either \cite{Brat03PRL,Bratnew} -- cf.
Fig. \ref{Fig1ab} (r.h.s.). The HSD calculations also demonstrate
that the 'partonic' Cronin effect plays a minor role at AGS and
SPS energies for the  parameter $T$.  The slope parameters from
$pp$ collisions (r.h.s. in Fig.  \ref{Fig1ab}) are seen to
increase smoothly with energy both in the experiment (full
squares) and in the transport calculations (full lines with open
circles) and are significantly lower than those from central Au+Au
reactions for $\sqrt{s} > 3.5$ GeV.

Thus the pressure generated by hadronic interactions in the transport
models above $\sim$ 5 A$\cdot$GeV is lower than observed in the
experimental data. This finding suggests that the additional pressure -
as expected from lattice QCD calculations at finite quark chemical
potential and temperature - might be generated by strong interactions in the
early pre-hadronic/partonic phase of central Au+Au (Pb+Pb) collisions.

\section{Charm signals of the QGP} \label{s3}

The microscopic HSD transport calculations (employed here) provide
a suitable space-time geometry of the nucleus-nucleus reaction and
a rather reliable estimate for the local energy densities
achieved.  The energy density $\varepsilon({\bf r};t)$ -- which is
identified with the matrix element $T^{00}({\bf r};t)$ of the
energy momentum tensor in the local rest frame at space-time
$({\bf r},t)$ -- reaches up to 30~GeV/fm$^3$ in a central Au+Au
collision at $\sqrt{s}$ = 200 GeV~\cite{Olena2}.

According to present knowledge the charmonium production in
heavy-ion collisions, {\it i.e.}  $c\bar{c}$ pairs,  occurs
exclusively at the initial stage of the reaction in primary
nucleon-nucleon collisions. The parametrizations of the total
charmonium cross sections ($i = \chi_c, J/\Psi, \Psi^\prime$) from
$NN$ collisions as a function of the invariant energy $\sqrt{s}$
used in this work are taken
from~\cite{Cass99,Cass00,brat03,Cass01}. We recall that (as in
Refs. \cite{brat03,Cass01,Geiss99,Cass97,CassKo}) the charm
degrees of freedom in the HSD approach are treated perturbatively
and that initial hard processes (such as $c\bar{c}$ or Drell-Yan
production from $NN$ collisions) are `precalculated' to achieve a
scaling of the inclusive cross section with the number of
projectile and target nucleons as $A_P \times A_T$ when
integrating over impact parameter. For fixed impact parameter $b$
the $c\bar{c}$ yield then scales with the number of binary hard
collisions $N_{bin}$ ({\it cf.} Fig. 8 in Ref.~\cite{Cass01}).

In the QGP `threshold scenario', e.g the geometrical Glauber model
of Blaizot et al.~\cite{Blaizot} as well as the percolation model
of Satz~\cite{Satzrev}, the QGP suppression `(i)' sets in rather
abruptly as soon as the energy density exceeds a threshold value
$\varepsilon_c$, which is a free parameter. This version of the
standard approach  is motivated by the idea that the charmonium
dissociation rate is drastically larger in a quark-gluon-plasma
(QGP)  than in a hadronic medium~\cite{Satzrev}.

On the other hand, the extra suppression of charmonia in the high
density phase of nucleus-nucleus collisions at SPS
energies~\cite{NA50aa,NA60} has been attributed to inelastic
comover scattering ({\it
cf.}~\cite{Cass99,Cass00,Cass97,Olena,Capella,Vogt99,Gersch,Kahana,Spieles,Gerland}
and Refs. therein) assuming that the corresponding $J/\Psi$-hadron
cross sections are in the order of a few
mb~\cite{Haglin,Konew,Sascha}. In these models `comovers' are
viewed not as asymptotic hadronic states in vacuum but rather as
hadronic correlators (essentially of vector meson type) that might
well survive at energy densities above 1 GeV/fm$^3$. Additionally,
alternative absorption mechanisms  might play a role such as gluon
scattering on color dipole states as suggested in
Refs.~\cite{Kojpsi,Rappnew,Blaschke1,Blaschke2} or charmonium
dissociation in the strong color fields of overlapping
strings~\cite{Geiss99}.

The explicit treatment of initial $c\bar{c}$ production by primary
nucleon-nucleon collisions and the implementation of the comover
model - involving a single matrix element $M_0$ fixed by the data
at SPS energies - as well as the QGP threshold scenario in HSD are
described in Refs.~\cite{Olena2,Olena} (see Fig.~1 of
Ref.~\cite{Olena} for the relevant cross sections). We recall that
the `threshold scenario' for charmonium dissociation is
implemented as follows: whenever the local energy density
$\varepsilon(x)$ is above a threshold value $\varepsilon_j$ (where
the index $j$ stands for $J/\Psi, \chi_c, \Psi^\prime$), the
charmonium is fully dissociated to $c + \bar{c}$. The default
threshold energy densities adopted are $\varepsilon_1 = 16$
GeV/fm$^3$ for $J/\Psi$, $\varepsilon_2 = 2$ GeV/fm$^3$ for
$\chi_c$, and $\varepsilon_3 =2 $ GeV/fm$^3$ for $ \Psi^\prime$.
Two more scenarios were implemented similarly to the `comover
suppression' and the `threshold melting' by adding the only
additional assumption -- that the comoving mesons (including the
$D$-mesons) exist only at energy densities below some energy
density $\epsilon _{cut}$, which is a free parameter. We use
$\epsilon _{cut}=1$~GeV/fm$^3$, {\it i.e.} of the order of the
critical energy density.

\begin{figure}%[!]
%\phantom{a}
%\vspace*{-0.5cm}
\centerline{ \psfig{figure=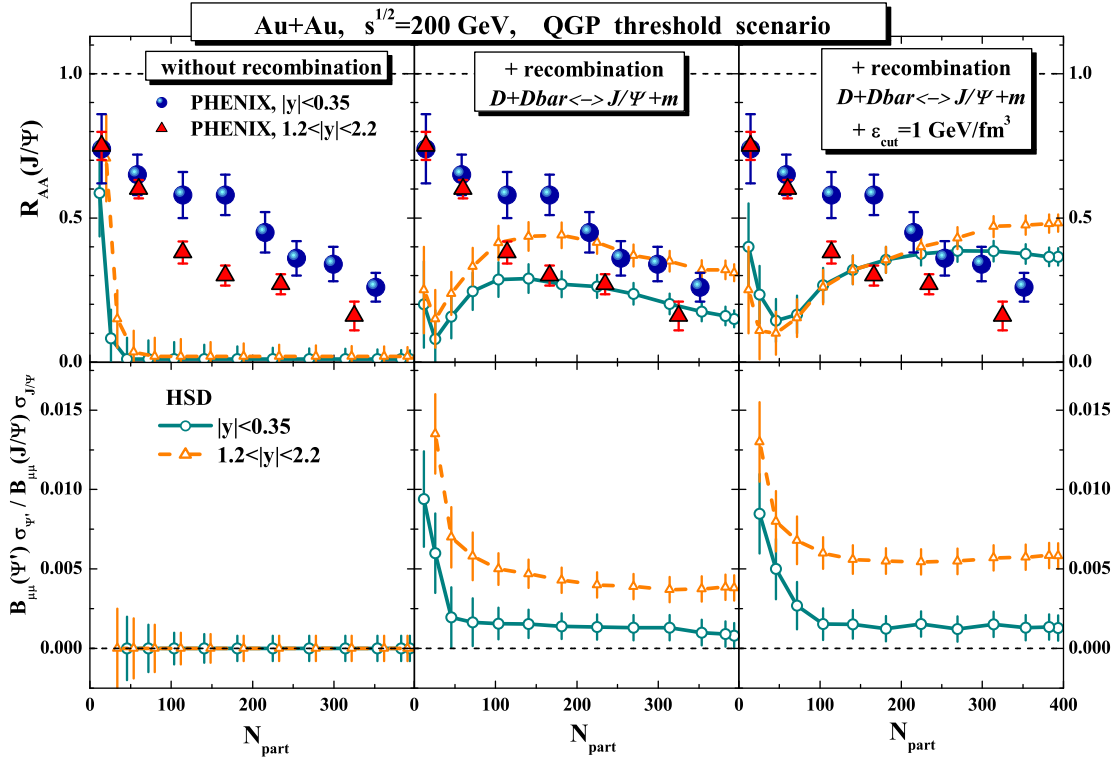,width=15cm}} \caption{The
$J/\Psi$ nuclear modification factor $R_{AA}$ for $Au+Au$
collisions at $\sqrt{s}=200$~GeV as a function of the number of
participants $N_{part}$ in comparison to the data from
\cite{PHENIX} for midrapidity (full circles) and forward rapidity
(full triangles). HSD results for the QGP `threshold melting'
scenarios are displayed in terms of the lower (green solid) lines
for midrapidity $J/\Psi$'s ($|y| \le 0.35$) and in terms of the
upper (orange dashed) lines for forward rapidity ($1.2 \le y \le
2.2$) within different recombination scenarios (see text). The
error bars on the theoretical results indicate the statistical
uncertainty due to the finite number of events in the HSD
calculations. Predictions for the ratio $B_{\mu \mu} (\Psi ')
\sigma _{\Psi'} / B_{\mu \mu} (J/\Psi) \sigma _{J/\Psi} $ as a
function of the number of participants $N_{part}$ are shown in the
lower set of plots. The figure is taken from~\cite{Olena2}.}
\label{RHICthreshold}
\end{figure}

\begin{figure}[t]
%\phantom{a}
%\vspace*{-0.5cm}
\centerline{\psfig{figure=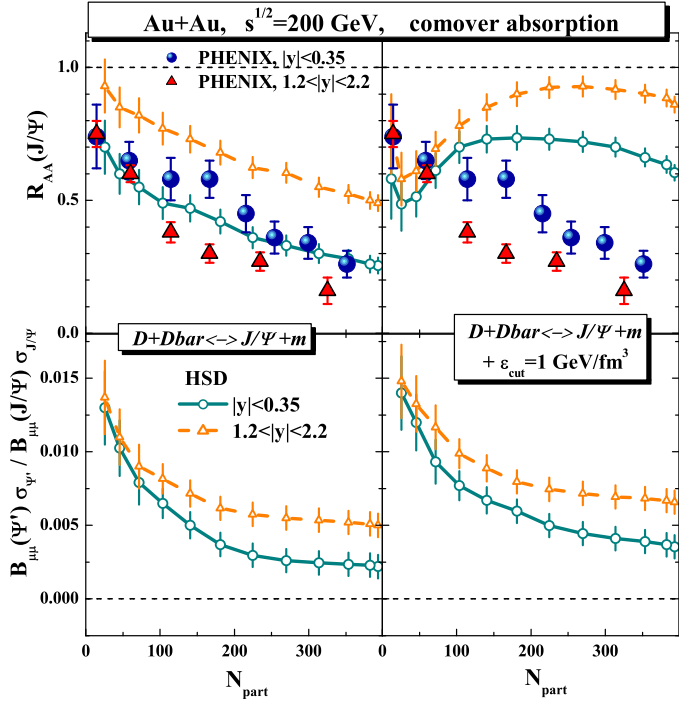,width=9cm}} \caption{Same as
Fig.~2 for the `comover absorption scenario' including the
charmonium reformation channels without cut in the energy density
(l.h.s.) and with a cut in the energy density $\epsilon _{cut} =
1$~GeV/fm$^3$ (see text for details). The figure is taken
from~\cite{Olena2}. } \label{RHICcomover}
\end{figure}

In the following, we compare our calculations to the experimental
data at the top RHIC energy of $\sqrt{s}=200$~GeV. We recall that
the experimentally measured nuclear modification factor $R_{AA}$
is given by,
\begin{equation}
R_{AA}=\frac{ d N (J/\Psi) _{AA} / d y  }{ N_{coll} \cdot d N
(J/\Psi) _{pp} / d y },
\end{equation}
where $d N (J/\Psi) _{AA} / d y $ denotes the final yield of
$J/\Psi$ in $A A$ collisions, $d N (J/\Psi) _{pp} / d y$ is the
yield in elementary $p p$ reactions and $N_{coll}$ is the number
of binary collisions.

Due to the very high initial energy densities reached
(corresponding to $T \geq 2 T_c$), in the threshold melting
scenario all initially created $J/\Psi$, $\Psi'$ and $\chi _c$
mesons melt. However, the PHENIX collaboration has found that at
least 20\%  of $J/\Psi$ do survive at RHIC~\cite{PHENIX}. Thus,
the importance of charmonium recreation is shown again. In HSD, we
account for $J/\Psi$ recreation via the $D  \bar D$ annihilation
processes as explained in detail in~\cite{Olena2,Olena}. Note that
in our approach, the cross sections of charmonium recreation in $D
+ \bar D \to J/\Psi + meson$ processes is fixed by detailed
balance from the comover absorption cross section $J/\Psi + meson
\to D + \bar D$. But even after both these processes are added to
the threshold melting mechanism, the centrality dependence of the
$R_{AA} (J/\Psi)$ cannot be reproduced in the `threshold melting'
scenario, especially for peripheral collisions (cf.
Fig.~\ref{RHICthreshold}). This holds for both possibilities: with
(r.h.s. of Fig.~\ref{RHICthreshold}) and without (center of
Fig.~\ref{RHICthreshold})  energy-density cut $\epsilon_{cut}$,
below which $D$-mesons and comovers do exist and can participate
in $D + \bar D \leftrightarrow J/\Psi + meson$ reactions.

Comover absorption scenarios give generally a correct dependence
of the yield on the centrality. If an existence of D-mesons at
energy densities above 1 GeV/fm$^3$ is assumed, the amplitude of
suppression of $J/\Psi$ at mid-rapidity is also well reproduced
(see the line for `comover without $\epsilon_{cut}$' scenario in
Fig.\ref{RHICcomover}, l.h.s.). Note that this line corresponds to
the prediction made in the HSD approach in~\cite{brat04}. On the
other hand, the rapidity dependence of the comover result is
wrong, both with and without $\epsilon _{cut}$. If hadronic
correlators exist only at $\epsilon < \epsilon _{cut}$, comover
absorption is insufficient to reproduce the $J/\Psi$ suppression
even at mid-rapidity (Fig.~\ref{RHICcomover}, r.h.s.). But its
contribution to the charmonium suppression is, nevertheless,
substantial. The difference between the theoretical curves marked
`comover + $\epsilon _{cut}$' and the data shows the maximum
possible suppression that can be attributed to a deconfined
medium.

We mention that there are also alternative explanations of the
experimental data for the anomalous $J/\Psi$ suppression in A+A
collisions: e.g. formation of charmonia only at the phase boundary
as advocated by Andronic {\it et al.}  \cite{Andronic} in the
statistical hadronization model.

\section{Summary}

Summarizing this contribution, I want to point out that strange
hadron production in central Au+Au (or Pb+Pb) collisions is quite
well described in the independent transport approaches HSD and
UrQMD \cite{Weber02}. The exception are the pion rapidity spectra
at the highest AGS energy and lower SPS energies, which are
overestimated by both models.  As a consequence the HSD and UrQMD
transport approaches underestimate the experimental maximum
('horn') of the $K^+/\pi^+$ ratio at $\sim$ 20-30 A$\cdot$GeV
\cite{Weber02}. The inverse slope parameters $T$ for $K^\pm$
mesons from the HSD and UrQMD transport models are practically
independent of system size from $pp$ up to central Pb+Pb
collisions and show only a slight increase with collision energy,
but no 'step' in the $K^\pm$ transverse momentum slopes as
suggested by Ga\'zdzicki and Gorenstein \cite{SMES} in 1999 and
found experimentally by the NA49 Collaboration. The rapid increase
of the inverse slope parameters of kaons for collisions of heavy
nuclei (Au+Au) found experimentally in the AGS energy range,
however, is not reproduced by both models (see Fig.~\ref{Fig1ab}).
Since the pion transverse mass spectra -- which are hardly
effected by collective flow  -- are described sufficiently well at
all bombarding energies \cite{Bratnew}, the failure has to be
attributed to a lack of pressure. I have argued - based on lattice
QCD calculations at finite temperature and baryon chemical
potential $\mu_B$ \cite{Karsch,Fodor} as well as the experimental
systematics in the chemical freeze-out parameters \cite{Cleymans}
- that this additional pressure should be generated in the early
phase of the collision, where the 'transverse' energy densities in
the transport approaches are higher than the critical energy
densities for a phase transition (or cross-over) to the QGP. The
interesting finding of the analysis is, that pre-hadronic degrees
of freedom might already play a substantial role in central Au+Au
collisions at AGS energies above $\sim$~5~$A\cdot$GeV.

The formation and suppression dynamics of $J/\Psi$, $\chi_c$ and
$\Psi^\prime$ mesons has, furthermore,  been studied within the
HSD transport approach for $Au+Au$ reactions from FAIR to top RHIC
energies of $\sqrt{s}$ = 200 GeV. It is found that both the
`comover absorption' and `threshold melting' concepts fail
severely at RHIC energies~\cite{Olena2} whereas both models
perform quite well at SPS energies. The failure of the 'hadronic
comover absorption' model goes in line with its underestimation of
the collective flow $v_2$ of leptons from open charm decay
\cite{brat05}. This suggests that the dynamics of $c, \bar{c}$
quarks at RHIC energies are dominated by strong
pre-hadronic/partonic interactions of charmonia with the medium in
a strong QGP (sQGP), which  cannot be modeled by `hadronic'
scattering or described appropriately by color screening alone.

The evidence for creating a 'new state of matter', the sQGP, in
Au+Au collisions at RHIC is overwhelming and is additionally
supported by the strong suppression of high $p_T$ hadrons and
'far-side' jets - both observations being insufficiently described
by string/hadronic models \cite{Ca1,Ca2} -  as well as the
quark-number scaling of elliptic flow $v_2(p_T)$. The question now
reads: what are the properties of the sQGP?

\vspace{0.5cm} \noindent {\bf Acknowledgements:}
I finally like to thank Mark Gorenstein for the numerous
discussions and fruitful collaboration over the last years.
Also a lot of thanks to my coauthors M. Bleicher, W. Cassing, O. Linnyk,
H. St\"ocker and H. Weber who contributed to the results presented
here.

%}end of \large


\begin{thebibliography}{99}

\bibitem{bjorken}
    J.D. Bjorken, Phys. Rev. D {\bf 27}, 140 (1983).
\bibitem{Karsch}
    F. Karsch {\it et al.}, Nucl. Phys. B {\bf 502}, 321 (2001).
\bibitem{exita}
    W. Cassing {\it et al.}, Nucl. Phys. A {\bf 674}, 249 (2000).
\bibitem{Fodor}
    Z. Fodor, S. D. Katz, and K. K. Szabo,
    Phys. Lett. B {\bf 568}, 73 (2003).
\bibitem{HORST}
    H. St\"ocker and W. Greiner, Phys. Rept. {\bf 137}, 277 (1986).
\bibitem{Geiss}
    J. Geiss {\it et al.}, Nucl. Phys. A {\bf 644}, 107 (1998).
\bibitem{Cass99}
    W. Cassing and E. L. Bratkovskaya, Phys. Rep. {\bf 308}, 65 (1999).
\bibitem{Sibbi}
	A. Sibirtsev and W. Cassing, Nucl. Phys. A {\bf 641}, 476 (1998).
\bibitem{Falter}
	T. Falter {\it et al.}, Phys. Lett. B {\bf 594}, 61 (2004);
	Phys. Rev. C {\bf 70}, 054609 (2004).
\bibitem{URQMD1}
    S.A. Bass {\it et al.}, Prog. Part. Nucl. Phys. {\bf 42}, 255 (1998).
\bibitem{URQMD2}
    M.~Bleicher {\it et al.},  J. Phys. G {\bf 25}, 1859 (1999).
\bibitem{Rafelski1}
    J.~Rafelski and B.~M\"uller, Phys. Rev. Lett. {\bf 48}, 1066 (1982).
\bibitem{SMES}
    M. Ga\'zdzicki and M. I. Gorenstein,
    Acta Phys. Polon. B {\bf 30}, 2705 (1999).
\bibitem{NA49_new}
     S. V.Afanasiev {\it et al.}, NA49 Collaboration,
     Phys. Rev. C {\bf 66}, 054902 (2002).
\bibitem{NA49_T}
    V. Friese {\it et al.}, NA49 Collaboration,
    J. Phys. G {\bf 30}, 119 (2004).
\bibitem{Weber02}
    H. Weber {\it et al.},
    {\em Phys. Rev.} C {\bf 67}, 014904 (2003).
\bibitem{Brat03PRL}
    E. L. Bratkovskaya {\it et al.}, \newblock { Phys. Rev. Lett.} {\bf 92}, 032302 (2004).
\bibitem{Bratnew}
    E. L. Bratkovskaya {\it et al.}, \newblock
    Phys. Rev. { C {\bf 69}}, 054907 (2004).
%------------------------------------------------------------
\bibitem{Olena2}
O.~Linnyk {\it et al.}, \newblock Phys. Rev. {\bf C76}, 041901
(2007).

\bibitem{Cass00}
W.~Cassing, E.~L. Bratkovskaya, and S.~Juchem,
\newblock Nucl. Phys. {\bf A674}, 249 (2000).

\bibitem{brat03}
E.~L. Bratkovskaya, W.~Cassing, and H.~{St\"ocker},
\newblock Phys. Rev. {\bf C67}, 054905 (2003).

\bibitem{Cass01}
W.~Cassing, E.~L. Bratkovskaya, and A.~Sibirtsev,
\newblock Nucl. Phys. {\bf A691}, 753 (2001).

\bibitem{Geiss99}
J.~Geiss {\it et al.}, \newblock Phys. Lett. {\bf B447}, 31
(1999).

\bibitem{Cass97}
W.~Cassing and E.~L. Bratkovskaya,
\newblock Nucl. Phys. {\bf A623}, 570 (1997).

\bibitem{CassKo}
W.~Cassing and C.~M. Ko,
\newblock Phys. Lett. {\bf B396}, 39 (1997).

\bibitem{Blaizot}
J.~P. Blaizot and J.~Y. Ollitrault,
\newblock Phys. Rev. Lett. {\bf 77}, 1703 (1996).

\bibitem{Satzrev}
H.~Satz,
\newblock J. Phys. {\bf G32}, R25 (2006).

\bibitem{NA50aa}
 M.~C. Abreu {\em et~al.},
\newblock Phys. Lett. {\bf B410}, 337 (1997);
{\bf B477}, 28 (2000);  {\bf B450}, 456 (1999).

\bibitem{NA60}
A.~Foerster {\em et~al.}, NA60 Collaboration,
\newblock J. Phys. {\bf G32}, S51 (2006).

\bibitem{Olena}
O.~Linnyk {\it et al.}, \newblock Nucl. Phys. {\bf A786}, 183
(2007).

\bibitem{Capella}
N.~Armesto and A.~Capella,
\newblock Phys. Lett. {\bf B430}, 23 (1998).

\bibitem{Vogt99}
R.~Vogt,
\newblock Phys. Rep. {\bf 310}, 197 (1999).

\bibitem{Gersch}
C.~Gerschel and J.~{H\"ufner},
\newblock Ann. Rev. Nucl. Part. Sci. {\bf 49}, 255 (1999).

\bibitem{Kahana}
D.~E. Kahana and S.~H. Kahana,
\newblock Prog. Part. Nucl. Phys. {\bf 42}, 269 (1999).

\bibitem{Spieles}
C.~Spieles {\em et~al.},
\newblock J. Phys. {\bf G25}, 2351 (1999),
\newblock Phys. Rev. {\bf C60}, 054901 (1999).

\bibitem{Gerland}
L.~Gerland {\em et~al.},
\newblock Nucl. Phys. {\bf A663}, 1019 (2000).

\bibitem{Haglin}
K.~L. Haglin,
\newblock Phys. Rev. {\bf C61}, 031903 (2000).

\bibitem{Konew}
Z.~Lin and C.~M. Ko,
\newblock Phys. Rev. {\bf C62}, 034903 (2000); J. Phys. {\bf G27}, 617 (2001).

\bibitem{Sascha}
A.~Sibirtsev, K.~Tsushima, and A.~W. Thomas,
\newblock Phys. Rev. {\bf C63}, 044906 (2001).

\bibitem{Kojpsi}
B.~Zhang, C.~M. Ko, B.-A. Li, Z.~Lin, and B.-H. Sa,
\newblock Phys. Rev. {\bf C62}, 054905 (2000).

\bibitem{Rappnew}
L.~Grandchamp, R.~Rapp,
\newblock Phys. Lett. {\bf B523}, 60 (2001),
\newblock Nucl. Phys. A {\bf 709}, 415 (2002).

\bibitem{Blaschke1}
D.~Blaschke, Y.~Kalinovsky, and V.~Yudichev,
\newblock Lect. Notes Phys. {\bf 647}, 366 (2004).

\bibitem{Blaschke2}
M.~Bedjidian {\em et~al.},
\newblock hep-ph/0311048.

\bibitem{PHENIX}
H.~{B\"usching} {\em et~al.}, PHENIX Collaboration,
\newblock Nucl. Phys. {\bf A774}, 103 (2006).

\bibitem{brat04}
E.~L. Bratkovskaya {\em et~al.},
\newblock Phys. Rev. {\bf C69}, 054903 (2004).

\bibitem{brat05}
E.~L. Bratkovskaya, W.~Cassing, H.~{St\" ocker}, and N.~Xu,
\newblock Phys. Rev. {\bf C71}, 044901 (2005).

\bibitem{Weber_stop02}
      H. Weber, E.L. Bratkovskaya and H. St\"ocker,
    { Phys. Lett.} B {\bf 545},  285 (2002).

\bibitem{Cleymans}
    J. Cleymans and K. Redlich, {Phys. Rev.} C {\bf 60},  054908 (1999).

\bibitem{Andronic}
  A.~Andronic {\em et~al.},
  Nucl.\ Phys.\  A {\bf 789}, 334 (2007) .

\bibitem{Ca1} { W. Cassing}, K. Gallmeister and C. Greiner,
Nucl. Phys A {\bf 735}, 277 (2004).

\bibitem{Ca2} K. Gallmeister and {W. Cassing},
Nucl. Phys A {\bf 748}, 241 (2005).

\end{thebibliography}
\end{document}